\documentclass[12pt]{article}
\newcommand{\text}{\rm}
\begin{document}

\title{\textbf{A gauge invariant and string independent fermion correlator in 
the Schwinger model}}
\author{D. G. Barci, L. E. Oxman and S. P. Sorella \and $\;$\vspace{4mm}\textbf{\ } \\
%EndAName
{\small {\textit{UERJ, Universidade do Estado do Rio de Janeiro}}} \\
{\small {\textit{Instituto de F\'{i}sica, Departamento de F\'{i}sica
Te\'{o}rica,}}} \\
{\small {\textit{\ Rua S\~{a}o Francisco Xavier 524, 20550-013 Maracan\~{a},
Rio de Janeiro, Brazil}}}\vspace{2mm}}
\maketitle

\begin{abstract}
We introduce a gauge invariant and string independent two-point fermion correlator 
which is analyzed in the context of the Schwinger model ($QED_2$). We also derive an effective infrared worldline action for this correlator, thus enabling the computation of its infrared behavior. Finally, we briefly discuss possible perspectives for the string independent correlator in the $QED_3$ effective models for the normal state of HTc superconductors.
\end{abstract}

\vfill\newpage\ \makeatother

\renewcommand{\theequation}{\thesection.\arabic{equation}}

\section{Introduction}

In the last years, a renewed interest in the study of quantum electrodynamics 
(QED) in lower dimensions ($2D$ and $1D$) is observed. One of the reasons is that 
several models of strongly correlated electrons are described in terms of fermions 
coupled to gauge fields. For instance, one promising model for the pseudo-gap phase 
of an underdopped cuprate system is based on a $2D$ gauge theory \cite{rw1,yft}. More precisely, 
this model describes the low energy fermionic excitations of a d-wave superconductor in 
such a way that the nodal quasi-particles are associated to massless Dirac fermions 
interacting with a gauge field, which represents the phase fluctuations of the superconductor 
gap. Vortex phase configurations play an important role for the dynamics of the model, after a coarse graining process they lead to a partition function coinciding with that of noncompact, parity conserving $QED_3$ \cite{ftv1}.

In the abovementioned systems, the absence of quasi-particles above the critical temperature $T_c$, observed in ARPES and tunneling experiments, suggests that the electron (hole) propagator has a Luttinger like behavior \cite{rw1},
\begin{equation}
\tilde{G}_{e}\sim \frac{\not\! p~~~}{|p|^{2-\eta}},
\label{mome}
\end{equation}
where $\eta > 0 $ is the anomalous dimension that controls the infrared (IR) behavior. Note that, in configuration space, a positive $\eta$ corresponds to an electron propagator with a decaying behavior which is faster than the noninteracting one ($\eta=0$). 

Wether the non-Fermi liquid behavior (\ref{mome}) is verified or not in the $QED_3$ model is  a nontrivial problem, since a detailed relationship among real electrons or holes and $QED_3$ fermions is unavailable. As a consequence, in the recent literature, a few proposals for the electron propagator have been introduced. In Refs.\cite{rw1,rw2}, the authors showed that an important requirement is that this propagator should be gauge invariant. In particular, they 
considered the correlator
\begin{equation}
G(x-y,\gamma)=\langle \psi(x)e^{ie\int_\gamma dx_\mu\, A_\mu} \bar{\psi}(y) \rangle, 
\label{gaugeinvariant}
\end{equation}
where $\gamma$ is a straight line running from $x$ to $y$.

A nonperturbative calculation of the anomalous exponent $\eta$ for the correlator (\ref{gaugeinvariant}) is a very hard task, while most perturbative approaches are very difficult to control, giving rise to some conflicting results about this parameter \cite{rw1}-\cite{gkr}. In fact, some of these calculations indicate a negative $\eta$, invalidating the interpretation of (\ref{gaugeinvariant}) as the propagator for electron modes with repulsive interactions \cite{rw2}-\cite{gkr}. This led to the search of other candidates to represent this propagator, such as the two-point correlation function of the nonlocal operator $\psi(x)\exp {i\partial^{-2} (\partial .A)}$ \cite{k1}, or to look for anomalous nonfermionic response functions \cite{fpst}. However, the identification of a proper correlator to represent the electron propagator still remains an open problem (see the discussion in \cite{gkr}).

In this work we shall consider the following gauge invariant, string independent two-point fermion correlator 
\begin{equation}
G_{inv}(x-y)=\int [d\gamma]\, \langle \psi(x)e^{ie\int_\gamma dx_\mu\, A_\mu} \bar{\psi}(y) \rangle,
\label{gammaintegration}
\end{equation}
where $\int[d\gamma]$ represents the average over classes of strings, running from $x$ to $y$, which are equivalent under reparametrizations. Our aim is to present a detailed computation of the IR behavior of (\ref{gammaintegration}) in the simpler and more tractable case of the Schwinger model \cite{sch}. This is motivated by the following reasons. 

Firstly, in $QED_2$ as well as in $QED_3$ the coupling constant is dimensionful, so that in both cases there is nothing to prevent the correlator (\ref{gaugeinvariant}) from a dependence on the shape of the string $\gamma$. Then, in both cases the string averaging in (\ref{gammaintegration}) will produce nontrivial modifications with respect to (\ref{gaugeinvariant}). 

Secondly, in Ref.\cite{k1}, the negative value of $\eta$ for the $QED_3$ string dependent correlator (\ref{gaugeinvariant}) was compared to the situation in $QED_2$, where this correlator also displays an anomalous negative exponent (see Eq.(\ref{IRp})).\footnote{Note however that the $QED_2$ correlator is not singular at $p=0$, because of the dynamical mass generation for the gauge field.} Then, by studying the effect the string averaging has on the IR behavior of the  string independent correlator in $1D$, we expect to shed some light on the behavior this correlator could possibly display in the physically relevant $2D$ case. 

In $1D$, unlike the $2D$ case, there is a great simplification as (\ref{gaugeinvariant}) can be computed exactly for an arbitrary string $\gamma$. In this way we can concentrate ourselves to obtain an effective IR string or ``worldline'' action for $\gamma$, and then to study the effect the worldline path integration has on the correlator.
In section \S 2, we review the calculation of the fermion two-point Green's function by means of 
the decoupling technique. In section \S 3, we compute the effective IR worldline action in 
$QED_2$. The IR behavior of the gauge invariant, string independent correlator is presented 
in section \S 4, while the worldline path integration is detailed in the appendix.
Section \S 5 is devoted to the presentation of our conclusions and to the discussion of possible perspectives for the case of $QED_3$.

\section{Two-point $QED_2$ fermion Green's function}

As a warm up, and in order to present our conventions, we review in this section the 
decoupling technique for the computation of the Schwinger model Green's function \cite{rscha,n},
\begin{equation}
G_{SM}(x-y)=\frac{1}{\cal N}\int [d\psi][d\bar{\psi}][dA]\,
\psi(x) \bar{\psi}(y) e^{ -\int d^2x'\, {\cal L}_{SM}},
\label{corrSM}
\end{equation}
\begin{equation}
{\cal L}_{SM}=-\bar{\psi}(i\not\!\partial+ e \not\! A )\psi + 
\frac{1}{4} F_{\mu \nu}F_{\mu \nu}.
\end{equation}
We work in euclidean space, and consider a representation where $\gamma_5=\sigma_3$. In the Lorentz gauge, the gauge field 
can be written according to $A_\mu=-(1/e)\varepsilon_{\mu \nu}\partial_\nu \phi$. The fermion and gauge degrees of freedom can be decoupled by means of a chiral change of variables, 
$\psi=e^{\gamma_5\phi}\chi$, $\bar{\psi}=\bar{\chi} e^{\gamma_5\phi}$, that is,
\begin{equation}
\psi=\left( \begin{array}{c}
\psi_+          \\ \psi_-
\end{array}\right)=\left( \begin{array}{c}
e^\phi \chi_+          \\ e^{-\phi}\chi_-
\end{array}\right)\makebox[.5in]{,} \bar{\psi}=\left( \begin{array}{cc}
\psi_-^{\dagger} & \psi_+^{\dagger} \end{array}\right)=\left( \begin{array}{cc}
e^{\phi}\chi_-^{\dagger} & e^{-\phi}\chi_+^{\dagger} \end{array}\right).
\end{equation}
In terms of the new variables we have
\begin{equation}
{\cal L}_{SM}=-\bar{\chi}i\not\!\partial \chi + \frac{1}{2e^2}\phi \partial^2 \partial^2 \phi,
\end{equation}
and the fermion correlator reads
\begin{equation}
G_{SM}(x-y)=\frac{1}{\cal N}\int [d\chi][d\bar{\chi}][d\phi]\,
e^{\gamma_5\phi(x)}\chi(x) \bar{\chi}(y)e^{\gamma_5\phi(y)} J_F(\phi) e^{ -\int d^2x'\, {\cal L}_{SM}},
\label{corrSMd}
\end{equation}
where $J_F$ is Fujikawa's anomalous Jacobian \cite{f}:
\begin{equation}
J_F(\phi)=e^{-\frac{1}{2\pi}\int d^2x'\, \partial_\mu \phi \partial_\mu \phi}.
\end{equation}
This corresponds to a regularization where the gauge symmetry 
\begin{equation}
\psi(x)\rightarrow e^{i\eta(x)}\psi(x)\makebox[.5in]{,}
A_\mu\rightarrow A_\mu + \frac{1}{e} \partial_\mu \eta
\end{equation}
is preserved at the quantum level.

The different components of the matrix in (\ref{corrSMd}), have the form
\begin{equation}
\frac{1}{\cal N}\int [d\chi][d\bar{\chi}]\, \chi_\pm (x)\chi^\dagger_\pm(y) e^{\int d^2x'\, \bar{\chi}{i\gamma_{\mu}\partial_{\mu}} \chi} \int [d\phi]\, e^{\pm\phi(x)\mp\phi(y)} e^{-\int d^2x'\, \frac{1}{2}\phi\hat{O}\phi},
\end{equation}
where
\begin{equation}
\hat{O}=\frac{\partial^2\partial^2}{e^2}-\frac{\partial^2}{\pi}.
\end{equation}
The diagonal components vanish, as they involve
$\chi_+ \chi^\dagger_-$ and $\chi_- \chi^\dagger_+$ correlations. On the other hand, the
$\phi$-factor associated with the off-diagonal components is the same for both  
$\chi_+ \chi^\dagger_+$ and $\chi_- \chi^\dagger_-$ correlations. Therefore, performing the gaussian path integral the well known result is obtained,
\begin{equation}
G_{SM}(x-y)=S_F(x-y)e^{\beta(x-y)},
\label{wk}
\end{equation}
where $S_F$ is the free massless Dirac propagator and
\begin{equation}
\beta(x-y)=\left[O^{-1}(0) -O^{-1}(x-y)\right]\makebox[.5in]{,}
\hat{O} O^{-1}(x)=\delta^{(2)}(x)
\end{equation}
\begin{equation}
O^{-1}(x)=\pi\left[ \Delta_{\frac{e^2}{\pi}}(x)-\Delta_0(x)\right]\makebox[.5in]{,}
(\partial^2-\mu^2)\Delta_{\mu^2}(x)=\delta^{(2)}(x).
\label{inverse}
\end{equation}
Note that $O^{-1}(0)$ can be absorved by means of a wave function renormalization. For large
$|x|$ the massive Green function decays exponentially, while $\Delta_0(x)=(1/2\pi)\ln |x|$. Therefore, the asymptotic behavior of the fermion correlator in the Schwinger model is 
(see Ref.\cite{st}),
\begin{equation}
G_{SM}(x)\sim \frac{\not\! x~~~}{|x|^{3/2}}.
\end{equation}

\section{Effective IR worldline action for $QED_2$}

Let us consider the gauge invariant, string dependent correlator (\ref{gaugeinvariant}) for the Schwinger model, that is,
\begin{equation}
G_{SM}(x-y,\gamma)=\frac{1}{\cal N}\int [d\psi][d\bar{\psi}][dA]\,
\psi(x)e^{ie\int_\gamma dx_\mu'\, A_\mu} \bar{\psi}(y) e^{-\int d^2x'\, {\cal L}_{SM}},
\label{corrSMinvg}
\end{equation}
where $\gamma$ is a smooth open string running from $x$ to $y$. Using decoupling techniques, this correlator has been computed for a straight line in Refs.\cite{sw,svz}.
Again, the only nonvanishing components in (\ref{corrSMinvg}) are the
off-diagonal ones. For instance, the component containing the $\psi_+ \psi^\dagger_+$ correlation function reads
\begin{equation}
\frac{1}{\cal N}\int [d\chi][d\bar{\chi}]\,\chi_+ (x)\chi^\dagger_+(y) e^{\int d^2x'\, \bar{\chi}i\gamma_{\mu}\partial_{\mu} \chi} \int [d\phi]\, e^{\int d^2x'\,[n-i\varepsilon_{\mu \nu}\partial_\mu j_\nu]\phi(x')} e^{-\int d^2x'\, \frac{1}{2}\phi\hat{O}\phi},
\label{++}
\end{equation}
where
\begin{equation}
n(x')=\delta^{(2)}(x'-x)-\delta^{(2)}(x'-y)\makebox[.5in]{,}
j_{\mu }(x')=\int_{\gamma }dx_\mu''\,\delta ^{(2)}(x'-x'')
\end{equation}
and ${\cal N}={\cal N}_\chi {\cal N}_\phi$, with
\begin{equation}
{\cal N}_\chi=\int [d\chi][d\bar{\chi}]\, e^{\int d^2x'\, \bar{\chi}i\not\!\partial \chi}
\makebox[.5in]{,}{\cal N}_\phi=\int [d\phi]\, e^{-\int d^2x'\, \frac{1}{2}\phi\hat{O}\phi}.
\end{equation}
The $\phi$-gaussian integral in (\ref{++}) is
\begin{equation}
{\cal N}_\phi \exp{I},
\label{gaussian}
\end{equation}
where
\begin{equation}
I=\frac{1}{2}\int d^2x' d^2x''\,[n(x')-i\varepsilon\partial' j ]
O^{-1}(x'-x'')[n(x'')-i\varepsilon\partial'' j ].
\label{expo}
\end{equation}
This exponent can be written as $I=I_{nn}+I_{nj}+I_{jj}$, where $I_{nj}$ collects the terms involving one $n$-field and one $j$-current, and $I_{nn}$ (resp. $I_{jj}$) is the term containing two $n$-fields (resp. two $j$-currents). In particular, $I_{nn}$ is the same exponent 
that appears in the two-point fermion Green's function (\ref{wk}), so that,
\begin{equation}
I_{nn}=\beta(x-y).
\end{equation}
The term containing two $j$-currents results
\begin{equation}
I_{jj}=\beta(x-y)- \Gamma_\gamma \makebox[.5in]{,}\Gamma_\gamma=-\frac{1}{2}e^2\int_\gamma \int_\gamma dx'_\mu dx''_\mu\, \Delta_{\frac{e^2}{\pi}}(x'-x'').
\label{jj}
\end{equation}
For the crossed term we have,
\begin{eqnarray}
I_{nj}=~~i\,\Omega_\gamma&=&-i\int_\gamma dx'_\mu\, \varepsilon_{\mu \nu}\partial_\nu' [O^{-1}(x-x')-O^{-1}(y-x')]\nonumber \\
&=&-i\int_\gamma dx'_\mu\, \varepsilon_{\mu \nu}\left(\frac{x'_\nu-x_\nu}{|x-x'|} \partial_{\cal A} O^{-1}(|x-x'|)-\frac{x'_\nu-y_\nu}{|y-x'|} \partial_{\cal A} O^{-1}(|y-x'|)\right),\nonumber \\
\label{crossed}
\end{eqnarray}
where we have used the property that $O^{-1}(x)$ is a function of ${\cal A}=|x|$.
When the points $x$ and $y$ are joined by a straight line, $\Omega_\gamma$ vanishes, as $dx'$, $x'-x$ and $x'-y$ are parallel along this curve.

Let us compute now the large distance approximation for the worldline action, that is, the string dependent part of (\ref{expo}). To this aim, we make use of the approximation $|y-x|>>1/m$ ($m^2=e^2 /\pi$) and we consider a smooth string $x(s)$, satisfying
\begin{equation}
\hat{a}\cdot\frac{dx}{ds}>0 \makebox[.5in]{,} \hat{a}=\frac{y-x}{|y-x|}.
\label{strings}
\end{equation}
We underline that the worldline action is reparametrization invariant, it only depends on the worldline's shape. For the class of strings (\ref{strings}) we can redefine the parameter $s \rightarrow s(\tau)$, according to
\begin{equation}
\frac{ds}{d\tau}=\left[ \hat{a} \cdot \frac{dx}{ds}\right]^{-1}.
\end{equation}
Of course, in the new parametrization we have
\begin{equation}
\hat{a} \cdot \frac{dx}{d\tau}=1,
\end{equation}
which implies
\begin{equation}
x(\tau)=\hat{a}\tau + x +\hat{n}\, \varepsilon(\tau)\makebox[.5in]{,} \hat{a} \cdot \hat{n}=0,
\label{param}
\end{equation}
where $\varepsilon$ gives the string transverse deformation with respect to the straight line.
As the string runs from $x$ to $y$, we must have
\begin{equation}
\tau \in [0, |y-x|]\makebox[.5in]{,} \varepsilon (0)=0~~,~~\varepsilon(|y-x|)=0.
\label{bcond}
\end{equation}
Then, we can write
\begin{equation}
\Gamma_\gamma=-\frac{1}{2}e^2\int_0^{|y-x|} d\tau \int_0^{|y-x|} d\tau'\, 
\frac{dx}{d\tau} \cdot \frac{dx}{d\tau'}\,
\Delta_{\frac{e^2}{\pi}}\left(|x(\tau')-x(\tau)|\right).
\label{int}
\end{equation}
This is a double-line integral with a kernel which is localized on a scale of the order of $1/m$. Therefore, for a fixed $\tau$, the leading contribution to $\Gamma_\gamma$ comes
from values of $|\tau'-\tau| < \xi$, where 
$1/m \sim \xi << |y-x|$. On this range, and for smooth strings ($\dot{\varepsilon}^2<<1$),
we can make use, in the integrand of (\ref{int}), of the approximations
\begin{equation}
|x(\tau')-x(\tau)|^2\approx \left(1+\dot{\varepsilon}^2\right)(\tau'-\tau)^2,
\end{equation}
and
\begin{equation}
\frac{dx}{d\tau} \cdot \frac{dx}{d\tau'}\approx \left( 1+ \dot{\varepsilon}^2\right),
\end{equation}
to obtain,
\begin{eqnarray}
\Gamma_\gamma &\approx & -\frac{1}{2}e^2\int_0^{|y-x|} d\tau \int_{\tau-\xi}^{\tau+\xi} d\tau' \, \left( 1+ \dot{\varepsilon}^2\right)
\Delta_{\frac{e^2}{\pi}}\left(\sqrt{1+\dot{\varepsilon}^2}|\tau'-\tau|  \right)\nonumber \\
&=& -\frac{1}{2}e^2\int_0^{|y-x|} d\tau \int_{-\xi}^{+\xi} du \, \left( 1+ \dot{\varepsilon}^2\right)
\Delta_{\frac{e^2}{\pi}}\left(\sqrt{1+\dot{\varepsilon}^2}|u|  \right).
\end{eqnarray}
The integral over $u$ can be estimated by taking the limit $\xi \to \infty$. Moreovere, by means of the change of variables $ \sqrt{1+\dot{\varepsilon}^2}u \rightarrow u$, we get 
\begin{eqnarray}
\Gamma_\gamma &\approx& -\frac{1}{2}e^2\int_{-\infty}^{+\infty} du \, 
\Delta_{\frac{e^2}{\pi}}\left(|u| \right)\int_0^{|y-x|} d\tau\, 
\sqrt{ 1+ \dot{\varepsilon}^2}\nonumber \\
&\approx& -\frac{1}{2}e^2\int_{-\infty}^{+\infty} du \, 
\Delta_{\frac{e^2}{\pi}}\left(|u| \right)\int_0^{|y-x|} d\tau\, 
\left( 1+ \frac{1}{2}\dot{\varepsilon}^2\right).
\end{eqnarray}
Summarizing, 
\begin{equation}
\Gamma_\gamma \approx \mu |y-x| + \int_0^{|y-x|} d\tau\, \frac{1}{2}\mu \dot{\varepsilon}^2,
\label{gammaction}
\end{equation}
where
\begin{equation}
\mu=-\frac{1}{2}e^2\int_{-\infty}^{+\infty} d\tau\, \Delta_{\frac{e^2}{\pi}}
(|\tau|)=\frac{e\sqrt{\pi}}{4}.
\end{equation}

Let us turn now to the analysis of the crossed term (\ref{crossed}). In general, using (\ref{inverse}), this term contains two contributions coming from the massive and massless Green's functions, respectively.
 
At large distances, or equivalently, when $e^2/\pi$ is large, the massive $\Delta$-functions in (\ref{crossed}) are well localized near the extrema $x$ or $y$ of the string $\gamma$.  Thus, as before, the contribution to (\ref{crossed}) coming from this small (almost straight) segments is suppressed. Then, the large distance behavior of $\Omega_\gamma$ comes from the massless Green function and reads
\begin{equation}
\Omega_\gamma\approx\frac{1}{2}\int_\gamma dx'_\mu\, \varepsilon_{\mu \nu}\left(\frac{x'_\nu-x_\nu}{|x'-x|^2} -\frac{x'_\nu-y_\nu}{|x'-y|^2} \right).
\end{equation}
Using $\varepsilon_{\mu \nu}\partial_\nu \ln |x|=\partial_\mu \arctan (x_0/x_1)$, we obtain
\begin{eqnarray}
\Omega_\gamma&\approx &\frac{1}{2}\left( \lim_{x'\to y}-\lim_{x'\to x}\right)[\arg (x'-x)-\arg (x'-y)]
\nonumber \\
& = &\frac{1}{2}\left[ \left( \arg (y-x) -\lim_{x'\to x}\arg (x'-x)\right)-
\left( \lim_{x'\to y}\arg (x'-y) - \arg (x-y)\right) \right],\nonumber \\
\end{eqnarray}
where the limits are performed along the string $\gamma$. We can also write
\begin{equation}
\Omega_\gamma \approx -\frac{1}{2}(\Theta_x +\Theta_y),
\label{omap}
\end{equation}
where $\Theta_x$ (resp. $\Theta_y$) is the angle relative to the direction of $\hat{a}$ of 
the vector tangent to the string at the point $x$ (resp. $y$). We note that the angle $\Theta$ must be considered as positive (resp. negative) when the tangent vector is obtained from an anti-clockwise (resp. clockwise) rotation of $\hat{a}$.

Summarizing the results we have obtained, (\ref{++}) is given by
\begin{equation}
\frac{1}{{\cal N}_\chi}\int [d\chi][d\bar{\chi}]\,\chi_+ (x)\chi^\dagger_+(y) e^{\int d^2x'\, \bar{\chi}i\gamma_{\mu}\partial_{\mu} \chi} \, \exp {I},
\label{chi1}
\end{equation}
and at large distances,
\begin{equation}
I=2\beta (x- y) - \Gamma_\gamma + i\,\Omega_\gamma\approx 2O^{-1}(0)+\ln |y-x|-\mu |y-x| - \int_0^{|y-x|} d\tau\, \frac{1}{2}\mu \dot{\varepsilon}^2
-\frac{i}{2}(\Theta_x +\Theta_y).
\label{I1}
\end{equation}

\section{The gauge invariant, string independent $QED_2$ correlator and its IR behavior}

Now, let us consider the gauge invariant, string independent correlation function
(\ref{gammaintegration}) for the Schwinger model. Using (\ref{corrSMinvg}), (\ref{chi1}) and (\ref{I1}), the off-diagonal component in (\ref{gammaintegration}), containing the $\psi_+ \psi^\dagger_+$ correlation function, reads
\begin{eqnarray}
&&\frac{1}{{\cal N}_\chi}\int [d\chi][d\bar{\chi}]\,\chi_+ (x)\chi^\dagger_+(y) e^{\int d^2x'\, \bar{\chi}i\not\!\partial \chi} \, \int [d\gamma]\,\exp {I}\nonumber \\
& = & \frac{1}{{\cal N}_\chi}\int [d\chi][d\bar{\chi}]\,\chi_+ (x)\chi^\dagger_+(y) e^{\int d^2x'\, \bar{\chi}i\not\!\partial \chi} \, \exp{2\beta (x- y) }
 \int [d\gamma]\,\exp {(- \Gamma_\gamma + i\,\Omega_\gamma)}.\nonumber \\
\label{offd}
\end{eqnarray}
It is interesting to note that for every path $\gamma$ running from $x$ to $y$, we have a path $\bar{\gamma}$ which is obtained by reflecting $\gamma$ with respect to the straight 
line joining $x$ and $y$. It is easy to see that $\Gamma_{\bar{\gamma}}=\Gamma_\gamma$, while 
$\Omega_{\bar{\gamma}}=-\Omega_\gamma$. Therefore, the $\gamma$, $\bar{\gamma}$ pair contribution to the path integral is
\begin{equation}
(\exp i\,\Omega_\gamma + \exp -i\,\Omega_\gamma)\exp -\Gamma_\gamma.
\label{pair}
\end{equation}
When the $\chi_- (x)$, $\chi^\dagger_-(y)$ correlation function is considered, we have to replace  $n\to -n$ in the equation (\ref{++}), which implies $\Omega_\gamma \to -\Omega_\gamma$. Then, from (\ref{pair}), the factor coming from the string path integral coincides for both correlation functions, and we can write
\begin{equation}
G_{inv}(x-y) = S_F(x-y)\exp{2\beta (x- y) }
\int [d\gamma]\,\exp {(- \Gamma_\gamma + i\,\Omega_\gamma)}.
\label{invs}
\end{equation}

Now, in order to obtain the infrared behavior of $G_{inv}$, we will consider the large distance approximation (\ref{I1}) and the path integral in (\ref{invs}) performed over the subset of smooth strings characterized by the condition (\ref{strings}). On this subset, each equivalence class can be represented by means of the parametrization (\ref{param}), that is, we will path integrate over the string transverse fluctuations $\varepsilon$. We note that in terms of $\varepsilon$, the 
approximated $\Omega_\gamma$ in (\ref{omap}) is, 
\begin{equation}
\Omega_\gamma \approx
\frac{1}{2}(\dot{\varepsilon}(0)+\dot{\varepsilon}(|y-x|)).
\end{equation}
After a field renormalization we get
\begin{eqnarray}
G_{inv}(x-y) &\sim &\frac{\not\! x-\not\! y}{|x-y|} \exp {(-\mu |x-y|)}\times K(|y-x|) \nonumber \\ 
K(|y-x|)&=& \int [d\varepsilon]\, \exp {\left( -\int_0^{|y-x|} d\tau\, \frac{1}{2}\mu \dot{\varepsilon}^2+ \frac{i}{2}[\dot{\varepsilon}(0)+\dot{\varepsilon}(|y-x|)]\right)},\nonumber\\
\label{invapprox}
\end{eqnarray}
where the $\varepsilon$ path-integral is performed with the boundary conditions given in (\ref{bcond}). Of course, the saddle point contribution ($\varepsilon \equiv 0$) in (\ref{invapprox}) corresponds to the asymptotic behavior of the string dependent correlator (\ref{corrSMinvg}) evaluated on a straight line,
\begin{equation}
G_{SM}(x,\gamma)\sim \frac{\not\! x}{|x|} \exp {(-\mu |x|)}.
\label{asyinv}
\end{equation}
However, the path integral over the string introduces in (\ref{invs}) a nontrivial modification with respect to the asymptotic behavior (\ref{asyinv}). In fact, the integration over the 
string transverse fluctuations in (\ref{invapprox}) is equivalent to a problem of quantum mechanics, that is, 
the evaluation of the propagator for a one dimensional ``free particle'' with position $\varepsilon$, plus perturbations at the boundaries. This propagator is computed in the appendix and reads,
\begin{equation}
K(|y-x|)=\frac{const}{|y-x|^{1/2}} \exp\left( \frac{1}{2\mu|y-x|}\right).
\label{K}
\end{equation}
Then, at large distances, the gauge invariant, string independent two-point correlation 
function receives a factor $|x|^{-1/2}$ coming from the factor $K$ in Eq.(\ref{K}), that is, 
\begin{equation}
G_{inv}(x) \sim \frac{\not\! x~~~}{|x|^{3/2}} \exp {(-\mu |x|)}.
\label{inv-asym}
\end{equation}

Note that the behaviors (\ref{asyinv}) and (\ref{inv-asym}), corresponding to 
the string dependent and string independent correlators, respectively, can be written as
\begin{equation}
G_{SM}(x,\gamma)\sim -\frac{\not\! \partial}{\mu} \exp {(-\mu |x|)}
\makebox[.5in]{,}
G_{inv}(x)\sim-\frac{\not\! \partial}{\mu} \left(\frac{\exp {(-\mu |x|)}}{|x|^{1/2}}\right),
\label{1}
\end{equation}
while
\begin{equation}
\exp {(-\mu |x|)}\sim \int d^2p\,\frac{e^{ipx}}{(p^2+\mu^2)^{3/2}}
\makebox[.5in]{,}
\frac{\exp {(-\mu |x|)}}{|x|^{1/2}}
\sim \int d^2p\,\frac{ e^{ipx}}{(p^2+\mu^2)}.
\end{equation}
As a consequence, we obtain the low-momentum behaviors,
\begin{equation}
\tilde{G}_{SM}(p,\gamma)\sim \frac{\not\! p}{(p^2+\mu^2)^{3/2}}
 \makebox[.5in]{,}
\tilde{G}_{inv}(p)\sim \frac{\not\! p}{(p^2+\mu^2)}.
\label{IRp}
\end{equation}
From the expressions (\ref{1}) and (\ref{IRp}) it is apparent that the string path integration  yields a different exponent with respect to the case where a fixed curve $\gamma$ is considered.

\section{Discussion}

In this work, we have computed the effective IR worldline action for the string independent correlator (\ref{gammaintegration}) in the context of $QED_2$. We have 
obtained the IR behavior for this correlator, showing that the integration over the string fluctuations leads to an infrared decay which is faster than that associated with the string dependent one. Of course, in $1D$, the dynamical mass generation for the gauge field prevents a singular behavior of the correlators in (\ref{IRp}) when the limit $p\to 0$ is considered.

In the physically interesting noncompact, parity conserving $QED_3$ models for the normal phase of HTc superconductors, such dynamical mass generation is absent (see \cite{ftv1} and references therein). 
Despite this difference, $QED_2$ and $QED_3$ share some similarities.

Both models display a negative anomalous exponent for the string dependent correlator (\ref{gaugeinvariant}). In particular, in $QED_3$, some authors \cite{rw2}-\cite{gkr} have obtained the Luttinger like behavior (\ref{mome}) with $\eta_0=-32/(3\pi^2N)$,  $N$ being the number of flavors.

Also, both models contain a dimensionful coupling constant, so that a nontrivial worldline action is expected in $QED_3$. In particular, based on the string independent $QED_2$ correlator in Eq.(\ref{1}), which contains an additional factor $|x|^{-1/2}$ coming from the string average, we expect that a similar effect could take place in the case of $QED_3$. This would imply a string independent $QED_3$ correlator with a power like decaying behavior which is faster than the string dependent one. If this is the case, depending on the number of flavors $N$, the positive correction to $\eta_0$ coming from the string average could give a final positive anomalous dimension.

This leads us to conjecture that the gauge invariant, string independent correlator (\ref{gammaintegration}) is a candidate to represent a sensible electron propagator in $QED_3$ models for HTc superconductivity.
 
In this regard, it would be very useful to obtain a reliable approximation scheme where these points could be checked, as well as to analyze the possibility of deriving the string averaged correlator by means of a careful coarse graining process in the underlying model for the phase fluctuations of the superconductor gap.

\section*{Acknowledgements}

The Conselho Nacional de Desenvolvimento Cient\'{\i }fico e Tecnol\'{o}gico
CNPq-Brazil, the Funda{\c{c}}{\~{a}}o de Amparo {\`{a}} Pesquisa do Estado
do Rio de Janeiro (Faperj) and the SR2-UERJ are acknowledged for the
financial support.

\section*{Appendix}

Here, for completness, we compute the path integral over the strings in the second line of (\ref{invapprox}). In general, we have
\begin{eqnarray}
\lefteqn{\int [d\varepsilon]\, \exp {\left( -\int_{\tau_i}^{\tau_f} d\tau\, 
\frac{1}{2}\mu \dot{\varepsilon}^2+ \frac{i\lambda}{2}[\dot{\varepsilon}(\tau_i)+\dot{\varepsilon}(\tau_f)]\right)}=}\nonumber \\ 
&&=const \int \prod_{k=1}^{n-1}d\varepsilon_k\, \exp{\left[-\frac{\mu}{2} \sum_{j=0}^{n-1}\frac{(\varepsilon_{j+1}-\varepsilon_j)^2}{\delta}+\frac{i\lambda}{2}\left(\frac{\varepsilon_1-\varepsilon_0}{\delta}+\frac{\varepsilon_n-\varepsilon_{n-1}}{\delta}\right)\right]},\nonumber \\
\label{pinte}
\end{eqnarray}
where $\varepsilon_0=\varepsilon_i$, $\varepsilon_n=\varepsilon_f$, and the limit $\delta\to 0$ 
($ n\delta=\tau_f-\tau_i$) is understood. We can also write,
\begin{eqnarray}
(\ref{pinte})&=&const \int \prod_{k=1}^{n-1}d\varepsilon_k \prod_{j=0}^{n-1}dp_j\, e^{-\frac{1}{2\mu}p_j^2\delta}
e^{i(\varepsilon_{j+1}-\varepsilon_j)p_j} 
e^{\frac{i\lambda}{2}\left(\frac{\varepsilon_1-\varepsilon_0}{\delta}+\frac{\varepsilon_n-\varepsilon_{n-1}}{\delta}\right)}\nonumber \\
&=&const \int \prod_{k=1}^{n-1}d\varepsilon_k \prod_{j=0}^{n-1}dp_j\, e^{-\frac{1}{2\mu}p_j^2\delta}
e^{-i\varepsilon_0(p_0+\frac{\lambda}{2\delta})}e^{i\varepsilon_n(p_{n-1}+\frac{\lambda}{2\delta})}\times \nonumber \\
&&\times e^{i\varepsilon_1(p_0-p_1+\frac{\lambda}{2\delta})}e^{i\varepsilon_2(p_1-p_2)}\dots
e^{i\varepsilon_{n-2}(p_{n-3}-p_{n-2})}e^{i\varepsilon_{n-1}(p_{n-2}-p_{n-1}-\frac{\lambda}{2\delta})}
\nonumber \\
&=&const \int \prod_{j=0}^{n-1}dp_j\, e^{-\frac{1}{2\mu}p_j^2\delta}
e^{-i\varepsilon_0(p_0+\frac{\lambda}{2\delta})}e^{i\varepsilon_n(p_{n-1}+\frac{\lambda}{2\delta})}\times \nonumber \\
&&\times \delta(p_0-p_1+\lambda/2\delta)\delta(p_1-p_2)\dots
\delta(p_{n-3}-p_{n-2})\delta(p_{n-2}-p_{n-1}-\lambda/2\delta).\nonumber
\end{eqnarray}
Now, we can change variables $p_0+\frac{\lambda}{2\delta}\to p_0$ and $p_{n-1}+\frac{\lambda}{2\delta}\to 
p_{n-1}$, and integrate the $\delta$-functions:
\begin{eqnarray}
(\ref{pinte})&=&const \int dp_0 dp_{n-1}\prod_{j=1}^{n-2}dp_j\, 
\delta(p_0-p_1)\dots \delta(p_{n-2}-p_{n-1})\times
\nonumber \\
&&\times\, e^{-\frac{1}{2\mu}p_j^2\delta}
e^{-\frac{1}{2\mu}(p_0-\frac{\lambda}{2\delta})^2\delta}
e^{-\frac{1}{2\mu}(p_{n-1}-\frac{\lambda}{2\delta})^2\delta}
e^{-i\varepsilon_0 p_0}e^{i\varepsilon_n p_{n-1}}\nonumber \\
%&=&const \int dp_0\, e^{-\frac{1}{2\mu}(p_0-\frac{\lambda}{2\delta})^2\delta}
%e^{-\frac{1}{2\mu}p_0^2\delta (n-2)}
%e^{-\frac{1}{2\mu}(p_0-\frac{\lambda}{2\delta})^2\delta}
%e^{-i\varepsilon_0 p_0}e^{i\varepsilon_n p_0}\nonumber \\
&=&const \int dp_0\, e^{-\frac{1}{\mu}(p_0-\frac{\lambda}{2\delta})^2\delta}
e^{-\frac{1}{2\mu}p_0^2 (n-2)\delta}
e^{ip_0(\varepsilon_n-\varepsilon_0)}\nonumber \\
&=&const \int dp_0\, e^{-\frac{1}{2\mu}p_0^2 n\delta+p_0[\frac{\lambda}{\mu}+i(\varepsilon_n-\varepsilon_0)]-\frac{1}{\mu\delta}(\lambda/2)^2},
\end{eqnarray}
and performing the quadratic integral, we get
\begin{equation}
(\ref{pinte})=\frac{const}{|\Delta\tau|^{1/2}} \exp {\frac{-\mu(\Delta\varepsilon)^2 }{2|
\Delta\tau|}}\exp\left( i\lambda \frac{\Delta\varepsilon}{\Delta\tau}+\frac{\lambda^2}{2\mu\Delta\tau}\right),
\end{equation}
where $\Delta\tau=\tau_f-\tau_i$, and $\Delta\varepsilon=\varepsilon_f-\varepsilon_i$.   
Of course, when $\lambda=0$ we have the usual free particle propagator. 
In the case we are interested (see Eq.(\ref{invapprox})) we have $\lambda=1$, $\tau_i=0$, $\tau_f=|y-x|$, and $\varepsilon(\tau_f)=0=\varepsilon(\tau_i)$, the path integral factor is real and coincides with the result showed in (\ref{K}).

\end{document}